\documentstyle{mn} 
\input psfig.tex 
 
\newif\ifAMStwofonts 
 
\ifoldfss 
  \ifCUPmtlplainloaded \else 
    \NewTextAlphabet{textbfit} {cmbxti10} {} 
    \NewTextAlphabet{textbfss} {cmssbx10} {} 
    \NewMathAlphabet{mathbfit} {cmbxti10} {} % for math mode 
    \NewMathAlphabet{mathbfss} {cmssbx10} {} %  "   "    " 
  \fi 
  \ifAMStwofonts 
    \ifCUPmtlplainloaded \else 
      \NewSymbolFont{upmath} {eurm10} 
      \NewSymbolFont{AMSa} {msam10} 
      \NewMathSymbol{\upi}     {0}{upmath}{19} 
      \NewMathSymbol{\umu}     {0}{upmath}{16} 
      \NewMathSymbol{\upartial}{0}{upmath}{40} 
      \NewMathSymbol{\leqslant}{3}{AMSa}{36} 
      \NewMathSymbol{\geqslant}{3}{AMSa}{3E}

      \let\leq=\leqslant \let\le=\leqslant 
      \let\geq=\geqslant  
    \fi 
  \fi 
\fi % End of OFSS 
 
\ifnfssone 
  \newmathalphabet{\mathit} 
  \addtoversion{normal}{\mathit}{cmr}{m}{it} 
  \addtoversion{bold}{\mathit}{cmr}{bx}{it} 
  \newmathalphabet{\mathbfit} % math mode version of \textbfit{..} 
  \addtoversion{normal}{\mathbfit}{cmr}{bx}{it} 
  \addtoversion{bold}{\mathbfit}{cmr}{bx}{it} 
  \newmathalphabet{\mathbfss} % math mode version of \textbfss{..} 
  \addtoversion{normal}{\mathbfss}{cmss}{bx}{n} 
  \addtoversion{bold}{\mathbfss}{cmss}{bx}{n} 
  \ifAMStwofonts 
    \ifCUPmtlplainloaded \else 
      \UseAMStwoboldmath 
      \makeatletter 
      \new@mathgroup\upmath@group 
      \define@mathgroup\mv@normal\upmath@group{eur}{m}{n} 
      \define@mathgroup\mv@bold\upmath@group{eur}{b}{n} 
      \edef\UPM{\hexnumber\upmath@group} 
      \new@mathgroup\amsa@group 
      \define@mathgroup\mv@normal\amsa@group{msa}{m}{n} 
      \define@mathgroup\mv@bold\amsa@group{msa}{m}{n} 
      \edef\AMSa{\hexnumber\amsa@group} 
      \makeatother 
      \mathchardef\upi="0\UPM19 
      \mathchardef\umu="0\UPM16 
      \mathchardef\upartial="0\UPM40 
      \mathchardef\leqslant="3\AMSa36 
      \mathchardef\geqslant="3\AMSa3E 

      \let\leq=\leqslant \let\le=\leqslant 
      \let\geq=\geqslant  
    \fi 
  \fi 
\fi % End of NFSS release 1 
 
\ifnfsstwo 
  \DeclareMathAlphabet{\mathbfit}{OT1}{cmr}{bx}{it} 
  \SetMathAlphabet\mathbfit{bold}{OT1}{cmr}{bx}{it} 
  \DeclareMathAlphabet{\mathbfss}{OT1}{cmss}{bx}{n} 
  \SetMathAlphabet\mathbfss{bold}{OT1}{cmss}{bx}{n} 
  \ifAMStwofonts 
    \ifCUPmtlplainloaded \else 
      \DeclareSymbolFont{UPM}{U}{eur}{m}{n} 
      \SetSymbolFont{UPM}{bold}{U}{eur}{b}{n} 
      \DeclareSymbolFont{AMSa}{U}{msa}{m}{n} 
      \DeclareMathSymbol{\upi}{0}{UPM}{"19} 
      \DeclareMathSymbol{\umu}{0}{UPM}{"16} 
      \DeclareMathSymbol{\upartial}{0}{UPM}{"40} 
      \DeclareMathSymbol{\leqslant}{3}{AMSa}{"36} 
      \DeclareMathSymbol{\geqslant}{3}{AMSa}{"3E} 

      \let\leq=\leqslant \let\le=\leqslant 
      \let\geq=\geqslant  
    \fi 
  \fi 
\fi % End of NFSS release 2 
 
\ifCUPmtlplainloaded \else 
  \ifAMStwofonts \else % If no AMS fonts 
    \def\upi{\pi} 
    \def\umu{\mu} 
    \def\upartial{\partial} 
  \fi 
\fi

\def\simlt{\lower.5ex\hbox{$\; \buildrel < \over \sim \;$}}
\def\simgt{\lower.5ex\hbox{$\; \buildrel > \over \sim \;$}}
\def\ms{M$_{\odot}$}

\begin{document}

\title{Chemo-spectrophotometric evolution of spiral galaxies: \\
     IV. Star formation efficiency and effective ages of spirals }

\author[S. Boissier, A. Boselli, N. Prantzos and G. Gavazzi]
       {S. Boissier$^1$, A. Boselli$^2$,  N. Prantzos$^1$ and G. Gavazzi$^3$  \\
1) Institut d'Astrophysique de Paris, 98bis, Bd. Arago, 75104 Paris \\
2) Laboratoire d'Astronomie Spatiale, Traverse du Siphon, F-13376 Marseille 
Cedex 12, France \\
3) Universit\`a degli Studi di Milano - Bicocca, Pizza dell'Ateneo Nuovo 1,
20126 Milano, Italy}

\date{ }

\pagerange{\pageref{firstpage}--\pageref{lastpage}}
\pubyear{2000}
\maketitle

\label{firstpage}

\begin{abstract}

We study the star formation history of normal spirals
by using a large and homogeneous data sample of local galaxies.
For our analysis we utilise detailed models of chemical and spectrophotometric
galactic evolution, calibrated on the Milky Way disc. We find that
star formation efficiency is independent of galactic mass, while
massive discs have, on average, lower gas fractions and are redder
than their low mass
counterparts; put together, these findings convincingly suggest that
massive spirals are older than low mass ones. We evaluate the effective ages 
of the galaxies of our sample and we find that massive spirals must be several
Gyr older than low mass ones. We also show that these galaxies (having 
rotational velocities in  the 80-400 km/s range)
cannot have suffered extensive mass losses, i.e. they cannot have 
lost during their lifetime an amount of mass much larger than their 
current content of gas+stars.

\end{abstract}


\section{Introduction}

The Star Formation Rate (SFR) is the most important and the less well understood
ingredient in studies of galaxy evolution. Despite more than 30 years of
observational and theoretical work, the Schmidt law still remains popular
among theoreticians and compatible with most available observations
(e.g. Kennicutt 1998a).

It is well known that there are systematic trends in the SF history of the
Hubble sequence: The ratio $\Psi/\langle\Psi\rangle$ of the current SFR 
$\Psi$ to the past
average one $\langle\Psi\rangle$ (integrated over the galaxy's age)
increases as one goes from early to late type galaxies, 
albeit with a large dispersion within each morphological type. 
The Hubble sequence  seems to be determined by the caracteristic 
timescale for star formation, with
early type galaxies forming their stars in shorter timescales than those of
late types. However, this simple description of the Hubble sequence fails to answer
two important, and probably related, questions:
i) what determines the caracteristic timescale for SF in galaxies? and ii)
what is the role (if any at all) of a galaxy's mass? 

Gavazzi et al. (1996) found an anti-correlation between the SFR per unit total mass
and galactic luminosity; part of this trend may reflect the aforementioned dependence
of SFR on Hubble type, but it may also be that this trend is fundamentally related
to the galactic mass. In a recent work Bell and de Jong (2000) use a large sample
of spiral galaxies with resolved optical and near-infrared photometry and
find that it is rather surface density that drives the star formation history of
galaxies, while mass is a less important parameter.

Photometric studies alone cannot lift the age-metallicity
degeneracy, namely the fact that young and metal-rich stellar populations may
be redder than old and metal-poor ones. Studies of the chemical aspects of
galaxy evolution (i.e. gas fractions, star formation rates, metal abundances)
can help to tackle the problem from a different angle, but they are also
limited by the unknown history of gaseous flows from and to the system.
A study combining elements from both photometric and chemical evolution offers
the best chances to understand this complex situation.

In this work we study the star formation history of spirals by using detailed
models of chemical and spectro-photometric galactic evolution. The numerical code
has been presented in detail elsewhere (Boissier and Prantzos 1999, hereafter 
Paper I) and is only briefly described in Sec. 2.
It has been successfully applied to the modelling of global properties of spirals
(Boissier and Prantzos 2000, hereafter Paper II), as well as to the corresponding
abundance and photometric gradients (Prantzos and Boissier 2000, herefter Paper 
III). Based on the recent work of Boselli et al. (2000), we use here a large and
homogeneous sample of data for normal spirals (presented in Sec. 3), ideally
suited for the purpose of this work. Our results and the comparison to the
observations are presented in Sec. 4. They suggest that galactic mass is the main
driver of galactic evolution, although local surface density may also play a role.
We argue that massive galaxies are, on average, older than less massive ones, based 
on the fact that star formation efficiencies seem to be
independent of galactic mass, while gas fractions are systematically low in massive
spirals. Based on the observed star formation efficiencies and gas fractions,
we derive in Sec. 5 corresponding galactic ages by using simple analytical 
models of galactic chemical evolution, taking into account gaseous flows. 
We find that low mass spirals must be several Gyr younger than massive ones.
Also, we find that spirals with rotational
velocities in the 80-400 km/s range must not have suffered extensive mass losses
during their past history. Our conclusions are summarized in Sec. 6.

\section{ Modelling the evolution of disc galaxies}

\subsection{The model and the Milky Way}

In Boissier and Prantzos (1999, paper I), we presented a model for the 
chemical and spectrophotometric  evolution of 
the Milky Way disc.
We recall here the main ingredients of the model.  

The disc is simulated as an ensemble of concentric, independently evolving rings, slowly
built-up by infall of primordial composition. For each ring
we solve the classical equations of chemical evolution (e.g. Pagel 1997)
without the assumption of Instantaneous Recycling Approximation. We use:
stellar lifetimes from Schaller et al. (1992);
the yields of Woosley and Weaver (1995) 
for massive stars and Renzini and Voli (1981) for intermediate mass stars; 
SNIa producing Fe with a rate given in Matteucci and Greggio (1986); and
the Initial Mass Function of Kroupa et al. (1993) between 0.1 \ms \ 
and 100 \ms. The adopted star formation rate (SFR) varies with gas 
surface density $\Sigma_{g}$ and radius $R$ as: 
\begin{equation}
\label{equa1}
 \Psi(t,R) \ = \  \alpha \  \Sigma_g(t,R)^{1.5} \ V(R) \ R^{-1}
\end{equation}
where $V(R)$ is the rotational velocity, assumed to be 220 km/s in the largest part
of the Milky Way disc.
This radial dependence of the SFR is suggested by
the theory of star formation induced by density waves in spiral
galaxies (e.g. Wyse and Silk 1989). 
The efficiency $\alpha$ of the SFR  (Eq. 1) is fixed
by the requirement that the observed local gas fraction, at R=8 kpc from the 
Galactic centre  ($\sigma_{gas}\sim$0.2), is reproduced at T=13.5 Gyr
(our adopted value for the age of the local disc).

The disc is built up by infall with a rate exponentially decreasing in
time, and a characteristic time-scale $\tau_{inf}(R)$ increasing with radius
(as to mimic the inside-out formation of the disc). 
The relation between $\tau_{inf}(R)$ and
$\Sigma(R)$ (total surface density) is shown on Fig. 1
and is {\it a posteriori} justified, since it is crucial in shaping the
various radial profiles (of gas, SFR, abundances, etc) of the disc, which 
compare favourably to observations.

The spectro-photometric evolution is followed in a self-consistent way,
with the metallicity dependent stellar tracks from the Geneva group 
(Schaller  et al. 1992,
Charbonnel et al. 1996) and stellar spectra from Lejeune et al. (1997).
Dust absorption is included according to the prescriptions of  
Guiderdoni et al. (1998) and assuming a ``sandwich''
configuration for the stars and dust layers.

It turns out that
the number of observables explained by the model is much
larger than the number of free parameters. In particular		
the model reproduces present day ``global'' properties 
(amounts of gas, stars, SFR, and supernova rates), as well as	
the current disc luminosities in various wavelength bands 
and the corresponding radial profiles of gas, stars, SFR and metal abundances;
moreover, the adopted inside-out star forming scheme leads to a 
scalelength of $\sim$4 kpc in the B-band and $\sim$2.6 kpc in the K-band, 
in agreement with observations (see paper I).

\subsection{Scaling relations}

For a simplified extension of the model to the the case of other  
disc galaxies we adopt the ``scaling properties''
derived by Mo, Mao and White (1998, hereafter MMW98) in the framework of
the Cold Dark Matter (CDM) scenario for galaxy formation.
For the details, the reader should refer to 
MMW98 and Boissier and Prantzos (2000, paper II).
Discs form inside non baryonic dark matter haloes of various masses and concentration
factors. Assuming constant disc to halo mass ratios (here taken to be $m_d$=0.05), 
discs are characterized by two parameters: 
$V_C$, the circular velocity measuring the
mass of the disc, and $\lambda$, the spin parameter measuring its angular momentum. 
A disc is described by  its scale-length $R_d$ and
its central surface density $\Sigma_0$, which  can be related to the
ones of our Galaxy (designated by $MW$) by: 
\begin{equation}
\label{ScalE1}
\frac{R_d}{R_{d,MW}}  \  = \  \frac{\lambda}{\lambda_{MW}} \ \frac{V_C}{V_{C,MW}}
\end{equation}
and
\begin{equation}
\label{ScalE2}
\frac{\Sigma_0}{\Sigma_{0,MW}}  \  = \
 \left(\frac{\lambda}{\lambda_{MW}}\right)^{-2}
 \ \frac{V_C}{V_{C,MW}}
\end{equation}
The total mass of the disc $M_d$ is proportional 
to $V_C^3$, but  independent of $\lambda$.
The $\lambda$ distribution deduced from numerical simulations (see MMW98 and references 
therein) presents a maximum at $\lambda \sim 0.04-0.05$ and extends from 
$\sim$0.01 to $\sim$0.20. Since the value of $\lambda_{MW}$ is not far from the peak value,
we explored here values of $\lambda$ in the range 1/3 $\lambda_{MW}$ to 3 $\lambda_{MW}$.
Assuming that $\lambda$ and $V_C$ are independent, we
constructed models with velocities in the observed range 80 to 360 km/s.
We calculated the velocity profile resulting from the disc plus an isothermal dark
halo, and  we used Equ. 1 to calculate self-consistently the SFR (with the same
coefficient $\alpha$, ``calibrated'' on the Milky Way).

At this point it should be noticed that our scaling relations (2 and 3) 
are based on the assumption that angular momentum is conserved during the 
evolution of the disk. Numerical hydrodynamical simulations of galaxy formation
(e.g. Navarro and Steinmetz 1999
and references therein) do not support this idea; indeed, it is found that
as the baryonic halo gas cools down and collapses to the disk it loses
most of its angular momentum. In those conditions, the final configuration of the
disk cannot be related in a simple way to the initial $\lambda$ of the halo.
However, such simulations lead , in general, to disk sizes much smaller 
than observed. It is possible that star formation and feedback are not properly
described in those simulations. For instance, Sommer-Larsen, 
Gelato and Vedel (1999) found that delaying the cooling of the gas reduces
the loss of angular momentum. Since the situation is not clear yet, we make here
the assumption of disk evolution at constant angular momentum, a posteriori
justified by the fact that the resulting disk sizes are in agreement with observations
(see Paper II). For a more detailed discussion of these issues, see also
Cole et al. (2000).

\subsection{Infall timescales}

\begin{figure}
\centerline{\psfig{file=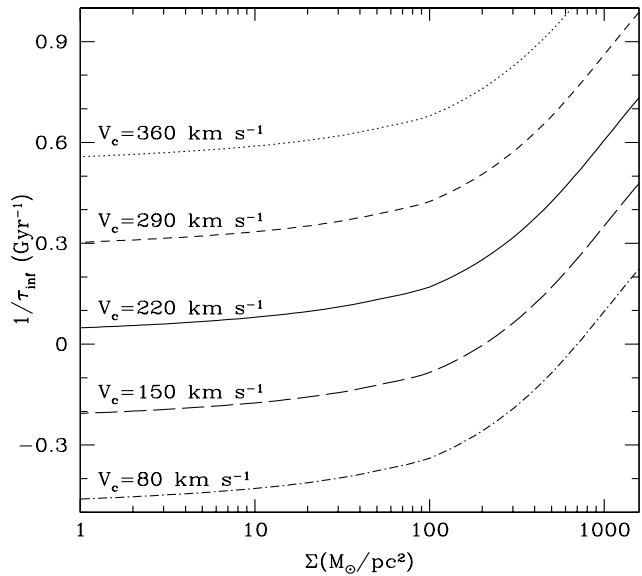,height=8cm,width=0.5\textwidth}} % tauBos.ps
\caption{\label{INFALL} Infall time-scale $\tau_{inf}$ as a function
of the local surface density $\Sigma$. The dependence on $\Sigma$
is calibrated on the Milky Way ({\it solid curve}, $V_C$=220 km/s). 
The dependence on
$V_C$ is chosen as to reproduce the properties of the homogeneous sample
of spiral galaxies presented in Sec. 3. When $\tau_{inf}<0$,
the infall is exponentially increasing with a time-scale
$| \tau_{inf} |$.
}
\end{figure}

The infall rate in our models is exponentially dependent on time,
with a time-scale $\tau_{inf}$ that depends on the surface
mass density and is calibrated on the Milky Way disc (see Fig. \ref{INFALL}); as
explained in paper II, we found  that a 
dependence on the total mass of the galaxy is necessary in order to
reproduce the observations of present-day discs.
We consider that all the galaxies started forming their stars 13.5 Gyr ago and that
the formation of their exponential discs (characterized
by $\Sigma_0$ and $R_d$) was completed at the present epoch.
Notice that the value of T=13.5 Gyr plays no essential 
role in the results of this work (values in the 10-15 Gyr range would lead
to quite similar conclusions); these results 
depend mainly on the infall timescales.
The more massive galaxies are characterized by shorter formation 
time-scales, while less massive galaxies are formed on longer ones 
(Fig. \ref {INFALL}).
This assumption turned out to be a crucial ingredient of our models,
allowing  to reproduce an impressive amount  of observed properties of spirals
that depend on mass (or  $V_C$): colours, gas fractions, abundances and
integrated spectra. 
Those observables  are thoroughly presented in paper II, while some of them 
are revisited  in this work.
In Prantzos and Boissier (2000, paper III in the series), we show that 
this model also reproduces fairly well the observed colour and abundance gradients
in spirals.

We notice that Bell and de Jong (2000) suggested recently that the observed 
colour gradients of disc galaxies correlate very well  with 
the local surface density, in the sense that inner and denser regions
are older. Our assumption about the infall timescale agrees, at least
qualitatively, with their findings.

We stress that
infall timescales are {\it inputs} to the model, and adjusted as to reproduce 
observations. Star formation timescales are {\it outputs} of the model (resulting
from the adopted prescriptions for infall {\it and} SFR) and are presented in Sec. 4.1.

\section {Comparison to observations}

\subsection{The observational sample}

The sample of galaxies analysed in this work, which has been extracted from
the large multifrequency database of nearby galaxies of Gavazzi and Boselli,
is extensively described in Boselli et al. (2000).
Here we give just a brief description of the sample selection criteria:
we refer the reader to Boselli et al. (2000) for the detailed references
on the data and on their analysis.\\

Galaxies analysed in this work are taken from the Zwicky catalogue 
(CGCG, Zwicky et al. 1961-1968)(m$_{pg}$ $\leq$ 
15.7). They are either late-type (type$>$S0a) 
members of 3 nearby (recession velocity c$z \le$ 8000 km s$^{-1}$) clusters 
(Cancer, A1367, Coma), or located in the relatively low-density regions of 
the Coma-A1367 supercluster 
(11$^h$30$^m$ $<$RA$<$ 13$^h$30$^m$; 
18$^o$$<$dec$<$32$^o$) as defined in Gavazzi et al. (1999a).
To extend the present study to lower luminosities,
we include in the sample the late-type Virgo cluster galaxies brighter 
than m$_{pg}$ $\leq$ 14.0 listed in the 
Virgo Cluster Catalogue as cluster members (VCC, Binggeli et al. 1985).
Furthermore VCC galaxies with
14.0 $\leq$ m$_{pg}$ $\leq$ 16.0 included in the ``ISO'' subsample 
described in Boselli et al. (1997a) and CGCG galaxies 
in the region 12$^h$ $<$RA$<$ 13$^h$; 
0$^o$$<$dec$<$18$^o$ but outside the VCC, are considered.

To avoid systematic environmental effects
we consider the subsample of late-type galaxies whose HI deficiency
(defined as $d$ = log($HI / \langle HI \rangle$),
the ratio of the HI mass to the 
average HI mass in isolated objects
of similar morphological type and linear size, see Haynes \& Giovanelli 1984)
is $d\geq$0.3, typical of unperturbed, isolated galaxies.
The final combined sample comprises 233, mainly ``normal''
galaxies.

We assume a distance of 17 Mpc for the members (and possible members) 
of Virgo cluster A, 22 Mpc for 
Virgo cluster B, 32 Mpc for
objects in the M and W clouds (see Gavazzi et al. 1999b).
Members of the Cancer, Coma and A1367 clusters are assumed at the 
distance of 62.6, 86.6 and 92 Mpc respectively.  
Isolated galaxies in the Coma supercluster are assumed 
at their redshift distance adopting $H_o$ = 75 km s$^{-1}$ Mpc$^{-1}$.

For the 233 optically selected galaxies,
data are available in several bands as follows:
100\% have HI (1420 MHz) and 99\% H band (1.65 $\mu$m) data, while 
a much coarser coverage 
exists in the UV (2000 \AA)(29 \%), CO (115 GHz)(38\%) and H$\alpha$ (6563 \AA)
(65\%), as shown in Table 1 and 2 of Boselli et al. (2000). The distribution
of the sample galaxies in the different morphological classes is given in 
Table 3 of Boselli et al. (2000).

As previously discussed, the present sample is optically selected and thus
can be biased against low surface brightness galaxies; the inclusion of
the Virgo cluster should in principle favor the presence of some
low surface brightness galaxies, not easily detectable at higher
distances. Being volume limited, the sample is not biased
toward bright, giant spirals, but it includes also dwarfs and compact
sources. Its completeness at different wavelengths makes this
a unique sample suitable for statistical analysis.

\subsection {Data analysis}

H$\alpha$ and UV fluxes, corrected for
extinction (and [NII] contamination) as described in Boselli et al. (2000),
are used to estimate star formation rates through population synthesis
models given in that work; a power-law IMF of slope -2.5 with a 
lower and upper mass cutoff of 0.1 \ms \ and  80 \ms, respectively, is adopted.
Its high mass part is quite similar to the one of the IMF of Kroupa et
al. (1993), adopted in our models.
Given the uncertainty in the UV and H$\alpha$ flux determination, in
the extinction correction and in the transformation of the corrected
fluxes into SFRs via population synthesis models, we estimate an 
uncertainty of a factor of $\sim$3 in the determination of the SFR.

The total gas content of the target galaxies, HI$+$H$_2$, has been
determined from HI and CO measurements. CO (at 2.6mm) fluxes have been 
transformed into H$_2$ masses assuming a standard CO to H$_2$
conversion factor of 1.0 10$^{20}$ mol cm$^{-2}$ 
(K km s$^{-1}$)$^{-1}$ (Digel et al. 1996). 
For galaxies with no CO measurement, we assume 
that the molecular hydrogen content is 10\% of the HI, as estimated 
from isolated spiral galaxies by Boselli et al. (1997b).
The total gas mass has  been corrected for He contribution by 30\%.
HI fluxes are transformed into neutral hydrogen masses
with an uncertainty of $\sim$ 10\%.  The average error on CO fluxes is 
$\sim$ 20\%; the error on the H$_2$ content, however, 
is significantly larger (and difficult to quantify) due
to the poorly known CO to H$_2$ conversion factor (see Boselli et al. 1997b).

Galaxy colours have been determined from broad band near-IR and optical 
photometry.
Near-IR (H band) images are available for 230 of the 233 sample galaxies, while
B images or aperture photometry for 214 objects. 
H and B magnitudes have been corrected for extinction as in 
Gavazzi \& Boselli (1996). No correction has been applied to galaxies of 
type later than  Scd. The estimated error on B and H magnitudes is 15 \%.

Rotational velocities have been determined from the HI line width at 21cm, and 
corrected for inclination as in Gavazzi (1987). To avoid large
systematic errors, we estimate rotational velocities only for galaxies 
with inclinations $>$ 30 deg and with the 21cm line width accurately determined
(double or single horned profile with high signal-to-noise). 
The uncertainty on the determination of the rotational velocity is $\sim$
15 km s$^{-1}$.

In summary, we have a homogeneous sample of 233 normal (non-perturbed) 
disc galaxies. For 96 of them, %for which
we evaluated all the quantities of interest in this work:
Blue magnitude $M_B$, total mass $M_T$, total gas mass $M_g$, 
star formation rate $\Psi$ and rotational
velocity $V_C$. We shall see below how our models fit those properties and what kind
of inferences can be made on the star formation history of those galaxies.

\subsection{Mass-driven colours and gas fractions}

\begin{figure}
\psfig{file=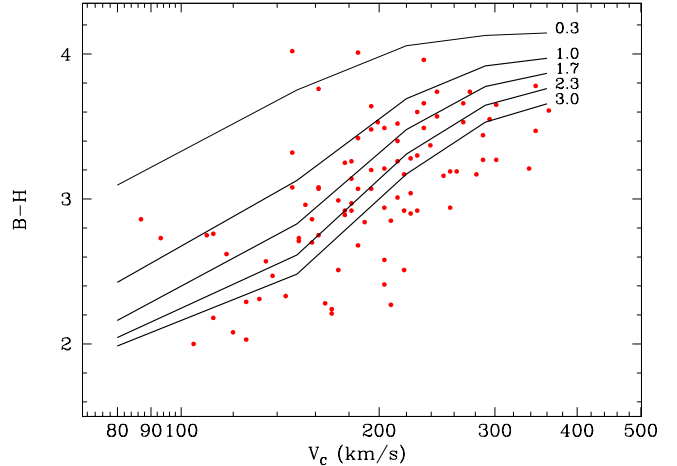,width=0.5\textwidth,angle=-90}
\caption{
Colour index $B-H$ vs rotational velocity $V_C$. Symbols are observations 
presented in Sec. 3.1. 
A clear correlation is obtained, with massive discs being redder than low mass
ones. The curves are the results of
our models and are labeled by the value of the ratio $\lambda/\lambda_{MW}$,
where $\lambda_{MW}$ is the spin parameter for the Milky Way disc.
The thin curve corresponds to very low values of $\lambda$, producing
very compact galaxies that look like bulges or ellipticals rather than spirals.
}
\end{figure}

In Fig. 2 we compare our model results to the colours $B-H$ of our  galaxy sample.
%In the upper panel we display all the galaxies of the sample, i.e.
%normal and perturbed spirals and normal and perturbed irregulars as well;
%there seems to be no corellation between colour and rotational velocity
%$V_C$  in that case. The situation changes when only normal spirals are taken
%into account (lower panel of Fig. 2): 
A clear correlation is obtained between
$B-H$ and $V_C$, with the  more massive discs being, on average, 
redder than their lower 
mass counterparts. Our models (solid curves) reproduce naturally the
observed correlation, since {\it by construction} the more massive discs
form their stars earlier (Fig. 1). The corresponding timescales for star
formation will be discussed in Sec. 4.1. The observed dispersion in the
lower panel of Fig. 2 can be accounted for by the range of $\lambda$ values
in our models, but only partially. Discs with lower $\lambda$ are more compact,
have higher central surface densities and evolve earlier than those with larger 
$\lambda$. Obviously, $\lambda$ values larger than 3 $\lambda_{MW}$
(the largest value used here), would lead to even smaller $B-H$ values
than the models displayed on Fig. 2, possibly accounting for the rest of the
scatter.

We notice that our models do take extinction by dust into account
(with the prescriptions presented in Sec. 2.1), and that extinction contributes
somewhat (by $\sim$0.5 mag) in redenning the most massive discs in our models, 
which have relatively large amounts of metals. 
However, we insist on the fact that it is {\it age}, not
extinction, which is mainly responsible for the trends of our models on Fig. 2.

In Fig. 3 we present two important pieces of data, on which the main argument 
of this paper is based. 
In the upper panel, we show the ratio $M_g/L_H$ (mass of gas to H-band luminosity)
vs rotational velocity $V_C$. Since $L_H$ reflects the mass of the stellar
population (Gavazzi et al. 1996), this ratio is a measure of the gas fraction of the system. Despite
the large scatter, it is clear that a correlation does exist, the more massive
galaxies being, in general, gas poor. Our model results can fairly well
describe the observations, accounting for both the slope and the dispersion of
the $M_g/L_H$ vs $V_C$ relation.

\begin{figure}
\psfig{file=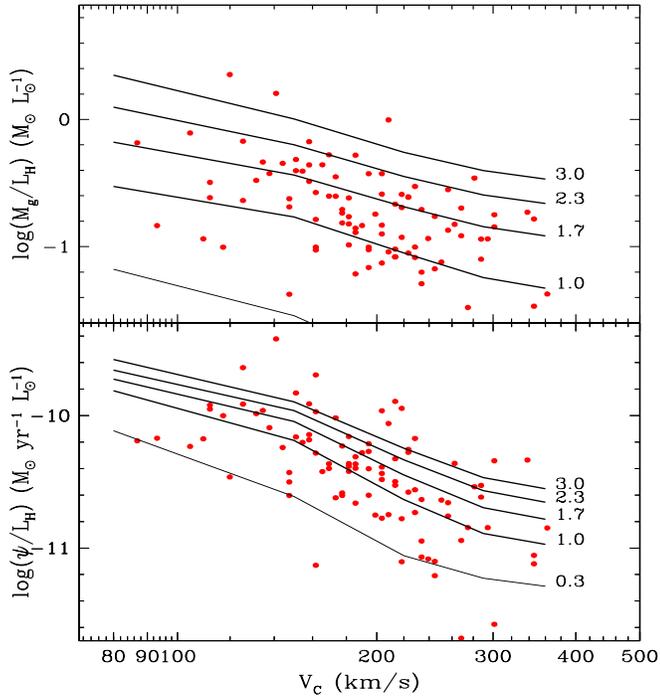,height=10cm,width=0.5\textwidth}
\caption{{\it Upper panel:} Ratio $M_g/L_H$
of gas mass to $H-$band luminosity vs rotational velocity $V_C$;
it is a measure of gas fraction. 
{\it Lower panel:} Ratio $\Psi/L_H$ of star formation rate to 
$H-$band luminosity vs. rotational velocity $V_C$; it is a measure
of the present SFR compared to the past average one. Massive discs
have smaller gas fractions and were more active in the past than their
lower mass counterparts. In both panels, our models (parametrised
with the spin parameter $\lambda/\lambda_{MW}$, where $\lambda_{MW}$ is the Milky Way 
value of $\lambda$) are in fair agreement with the data.
}
\end{figure}

In the lower panel of Fig. 3 we plot the ratio $\Psi/L_H$ (current SFR to
$H-$band luminosity) vs. $V_C$. As discussed in several places (e.g. Kennicutt 1998a),
this ratio is a measure of the parameter $b = \Psi/\langle\Psi\rangle$, the ratio
of the current SFR to the past average one. Massive galaxies display
smaller  $\Psi/L_H$, and thus lower $b$ values, than their low mass counterparts.
This means that they formed their stars at much higher rates in the past.
Again, our results compare fairly well with the data, concerning both the slope 
and the scatter of the correlation. For a given $V_C$, large $\lambda$ 
discs have larger gas fractions today and were less active in the past than their
lower $\lambda$ counterparts.
As we shall see in Sec. 4.3, and discuss in Sec. 5,
when the two panels of Fig. 3 are combined, they suggest quite convincingly
that low mass spirals are on average younger than massive ones.

At this point, we notice that Gavazzi and Scodeggio (1996)
already advanced a similar hypothesis to explain the observed colors of
galaxies as a function of their H-band luminosity (a mesure of their 
dynamical mass according to them).  Gavazzi et al. (1996) 
came to  similar conclusions and suggested that mass is the main parameter of 
galaxy evolution, on the basis of multiwavelength
observations concerning  a variety of disc properties
(colours, gas content, star formation rate, radius,
surface brightness).

\section{Star formation histories in discs}

\subsection{Formation time-scales}

\begin{figure}
\psfig{file=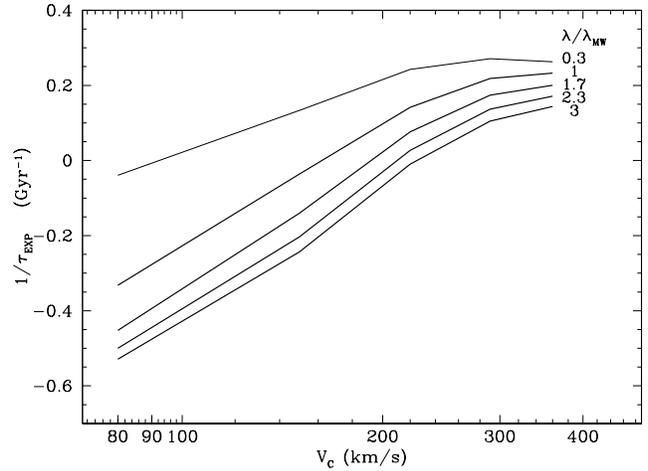,width=0.5\textwidth,angle=-90} 
\caption{\label{TIMES} 
Time-scale of an exponential fit to the star
formation rate history of our model galaxies; $\tau_{EXP}^{-1}$ (Gyr$^{-1}$) is plotted
as a function of disc rotational velocity $V_C$ and parametrised 
with the spin parameter $\lambda/\lambda_{MW}$. For  low values of $V_C$, $\tau_{EXP}$
is negative, i.e. the star formation rate is  increasing with time. 
$\tau_{EXP}^{-1}$ is larger for massive galaxies, which  form their stars earlier.}
\end{figure}

The Hubble sequence of galaxies is usually interpreted in terms of different
star formation timescales (e.g. Kennicutt 1998a), although such an interpretation
leaves unclear the role of the galaxy's mass (Prantzos 2000). In order to put
our results in that context, we performed an exponential fit to the star formation 
histories of our model galaxies (excluding the first 2 Gyr, where such a  fit turns
out to be inadequate). The resulting timescales are shown in Fig. \ref{TIMES}. Notice that
in some models the SFR is continuously increasing with time, resulting in negative
caracteristic timescales. For that reason
we present $\tau_{EXP}^{-1}$, which has the advantage of varying
continuously when going from a SFR increasing in time to one decreasing with time.
For an exponential star formation
rate, $\Psi \propto exp^{-t/\tau_{EXP}}$,
$\tau_{EXP}^{-1}$ is equal to $-\dot{\Psi}/\Psi$ (where $\dot{\Psi}=d\Psi/dt$) 
and can be considered
as  the normalized rate of change of  the SFR. 
Obviously, the larger is  $\tau_{EXP}^{-1}$,
the earlier the galaxy forms its stars.

As can be seen on Fig. 4, massive galaxies in our models
form their stars on shorter timescales
than their lower mass counterparts. This general trend is somewhat modulated
by the spin parameter $\lambda$: discs with smaller $\lambda$ (i.e. more compact)
have larger  $\tau_{EXP}^{-1}$ than discs with larger $\lambda$ of similar
rotational velocity. The most massive discs of our simulations ($V_C$=360 km/s)
have decreasing SFR with $\tau_{EXP}\sim$4-5 Gyr. Discs with $V_C\sim$200 km/s
have long timescales, of the order of $\sim$10 Gyr, i.e. essentially constant
SFR. Finally, low mass discs ( $V_C\sim$100 km/s) have SFR increasing in time, 
with caracteristic timescales $\tau_{EXP}^{-1}\sim$-0.2 to -0.5 Gyr$^{-1}$.

\subsection{The evolution of the star formation efficiency}

What is the reason for the vastly different SF timescales obtained in our models
as a function of $V_C$? Is the overall SF efficiency $\epsilon = \Psi/M_g$ 
directly affected by the mass of the galaxy? An inspection of Eq. 1
suggests that this cannot be the reason. Indeed, at the caracteristic radius
of the disc $R_d$, the local efficiency is:
\begin{equation}
\epsilon(R_d) \ = \ \Psi(R_d)/\Sigma_g(R_d) \ 
= \alpha \ \Sigma_g^{0.5} \ V \ R_d^{-1} \ \propto \ \lambda^{-2} \ V^{0.5}
\end{equation}
since $R_d \propto \lambda \ V$ from Eq. 2 and $ \Sigma(R_d) \propto \Sigma_0 \propto
\lambda^{-2} V$ from Eq. 3. In other terms, the SF efficiency at $R_d$ 
varies very little with $V_C$ and depends much more
on $\lambda$ than on $V_C$. 
Since the value of any intensive quantity (like SF efficiency) at
$R_d$ is typical of the whole disc, it is expected that the global SF efficiency
also depends little on $V_C$.
In order to show this quantitatively, we plot in Fig. 5
the evolution of the global SF efficiency $\epsilon$ of our models 
as a function of time. We show the
results for 3 values of $\lambda$ (1/3, 1 and 3 times $\lambda_{MW}$)
and three values of $V_C$ (80, 220 and 360 km/s, respectively).
We compare our results at T=13.5 Gyr to estimates of $\epsilon = \Psi/M_g$ in 
our sample of nearby spirals (presented in Sec. 4.3) . 
The following points should be noted concerning our models:

\begin{figure}
\psfig{file=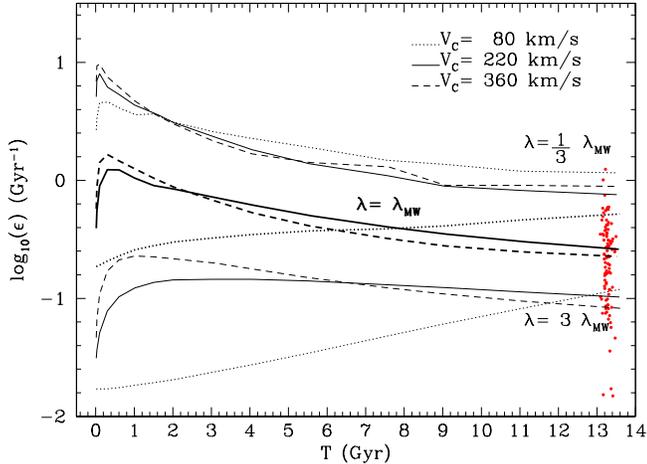,width=0.5\textwidth,angle=-90}
\caption{Evolution of the Star Formation Efficiency $\epsilon$
$(\Psi/M_{g})$ as a function of time for three families 
of models labeled by their $\lambda$
values (1/3, 1, and 3 times $\lambda_{MW}$). 
Low $\lambda$ values correspond to  more compact
galaxies, while values larger than  3 $\lambda_{MW}$ lead to
Low Surface Brightness Galaxies.
For each $\lambda$, three curves are shown corresponding to $V_C$=80, 220 and 360
km/s, respectively. The SF efficiency  
$\epsilon$ depends more on $\lambda$ than on $V_C$. At T=13.5 Gyr, the 
present-day observations (Sec. 4.3) are shown; an artificial small age
spread is introduced. Our  grid of
models at T=13.5 Gyr covers well the observed   $\epsilon$ values.
}
\end{figure}

i) The efficiency $\epsilon$ depends very little on $V_C$, especially during the
last half of the history of the galaxy.

ii) $\epsilon$ is mainly determined by $\lambda$: ``compact'' galaxies have higher
efficiency because of their smaller size ($\Psi \ \propto \ 1/R $,
for a given $V_C$) and larger gas surface
density at the caracteristic radius $R_d$. 
At T=13.5 Gyr, the variation of our $\epsilon$ values
due to   $\lambda$ can fully account for the dispersion
in $\epsilon$ measured in nearby spirals.

iii) The SF efficiency  in compact galaxies (low $\lambda$) 
presents a peak at early times and then
decreases. This happens because the star formation migrates to outer regions (due to the
inside-out disc formation scheme) 
where the local SF efficiency
is lower (because of lower surface densities and of the 1/R factor).
In more extended galaxies (larger $\lambda$)  
$\epsilon$ does not present such a decrease because local
properties vary little with radius.

iv) For the lowest disc velocities $V_C$, $\epsilon$ may increase with time. This is due to
the adopted form of infall: the gas surface density increases considerably 
when the gas arrives finally in the disc (which may take a very long time in the case
of the largest $\lambda$ values and lowest $V_C$ values, see Fig. 4).

\begin{figure*} 
\psfig{file=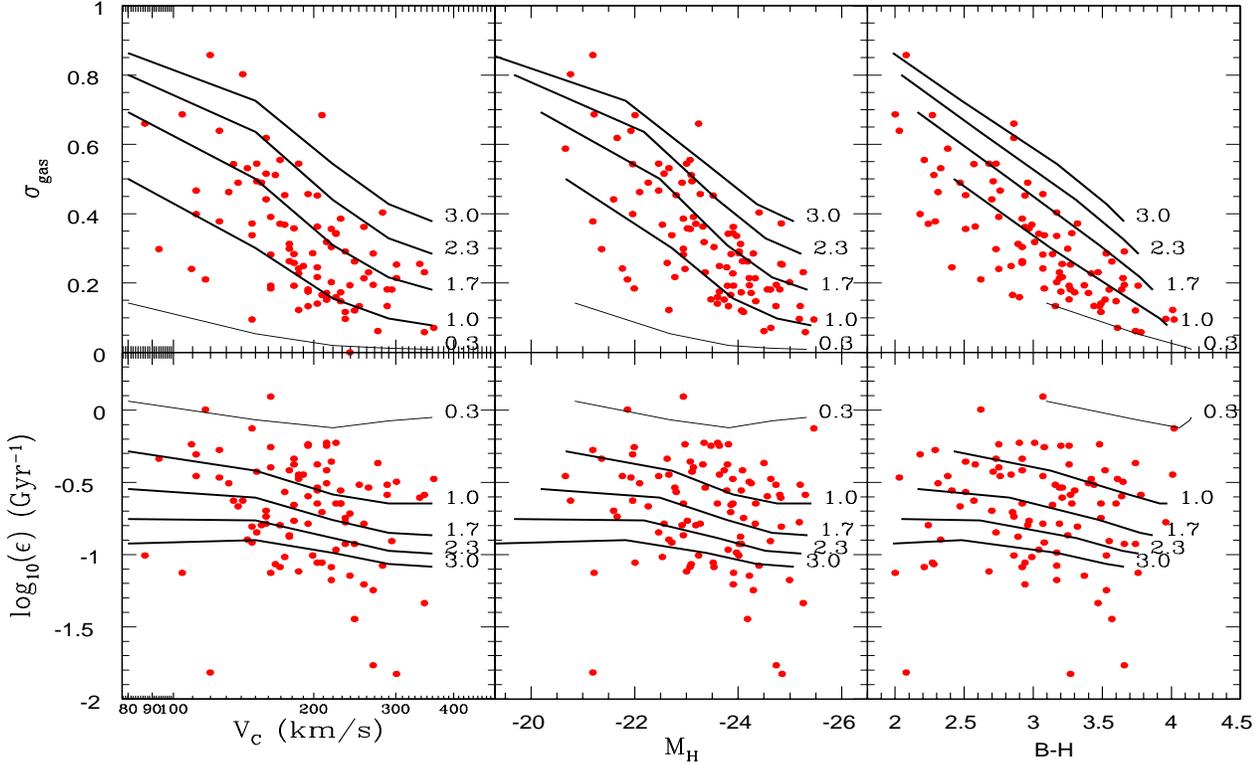,height=11cm,width=\textwidth,angle=-90} %gfeffNU6.ps
\caption{\label{GEFF} Gas fraction ($\sigma_{gas}=M_{g}/M_{T}$,
{\it upper panels}) and global 
Star Formation Efficiency ($\epsilon=\Psi/M_{g}$, {\it lower panels}) 
vs circular velocity $V_C$ ({\it left panels}), H-band
magnitude ({\it middle panels}) 
and colour index B-H ({\it right panels}). 
Dots represent our data (Sec. 4.3).
The results of our models (Sec. 4.2) are shown by curves, parametrised by the
the ratio $\lambda$/$\lambda_{MW}$, where $\lambda_{MW}$ is the
Milky Way spin parameter. The gas fraction decreases, on average with
galaxy's mass, luminosity and colour index, while the
global SF efficiency $\epsilon$ does not depend on those parameters.
Our models account fairly well for these data.
We note that $\epsilon_{model}$ is a function of
$\lambda$ and is greater for compact galaxies. 
}
\end{figure*}

The main point of this section is that in our models the SF efficiency $\epsilon$ does
not depend directly on the mass of the galaxy. The range of $\epsilon$ 
span by our models during galactic evolution is due to $\lambda$ and $\tau_{INF}$,
not to $V_C$; the most important of the two parameters is $\tau_{INF}$. Of course,
$\tau_{INF}$ is adjusted to $V_C$, so that the observations
of Fig. 2 and 3 (and many others, presented in Paper II) are reproduced. This is
a crucial ingredient for the success of our models, and we discuss its implications
in Sec. 5.

\subsection{Gas fraction and star formation efficiency}

The analysis of Sec. 4.1 and Sec. 4.2 lead to an important conclusion: the success 
of our models is to be interpreted in terms of mass-dependent SF timescales;
but this is not due to any explicit dependence of SF {\it efficiency}
on galaxy mass. Indeed, the SF efficiency of our models is virtually
independent of mass during most of galactic history, and in particular at the
present time. Is this supported by observations?

In Fig. 6 we display our data of Fig. 3, this time in a more ``physical''
presentation, appropriate for a quantitative discussion  (see Sec. 5).
The gas fractions $\sigma_{gas}$ (upper panels) and
SF efficiency $\epsilon$ (lower panels) are presented as a function of rotational
velocity $V_C$ (left panels), H-magnitude (middle panels) and
$B-H$ colour (right panels).

The gas fraction $\sigma_{gas}=M_g/M_T$ (where the total mass 
$M_T=M_*+M_g$ is the mass of stars + gas)
is obtained by converting the $L_H$ luminosity to star mass $M_*$ through 
the mass to light ratio $M_*/L_H$ obtained in our models. 
This value is in the 0.3-0.6 range in solar units ($M_*/L_H\sim$0.35 for
$V_C$=80 km/s, $\sim$0.47 for $V_C$ 220 km/s and $\sim$0.56 
for $V_C$=360 km/s).
Despite the uncertainties, a clear trend is present in the upper panel:
the more massive, luminous and red a galaxy is, the smaller is its gas fraction.
Notice that a similar trend is obtained by McGaugh and de Blok (1997) and
Bell and de Jong (2000).
Our model results are in excellent agreement with the data, although we have some
difficulty in reproducing blue and  gas poor discs (with $B-H<$2.5 and 
$\sigma_{gas}<$0.4).

The SF efficiency of our observed galaxies is obtained by simply
dividing the SFR $\Psi$ with the gas mass $M_g$.
We notice that
the uncertainties in  deriving $\epsilon$ are rather large: those concerning 
the SFR $\Psi$ are quite large (a factor of $\sim$3), whereas those of
the gas mass at least of $\pm$20 \%. As a result, the observationally derived
 scatter of  $\epsilon$, as appears on  Figs. 5 and 6, 
is certainly larger that the real one. It is clear, however, that
the SF efficiency does not seem to be correlated with the mass or the colours
of spirals. The scatter in the observed values of
$\epsilon$ (a factor of $\sim$10), should be compared to the 
range of $\sim$60 span by galaxy mass (since the mass of the disc 
$m_d \propto V_C^3$) or the range of five magnitudes in H-luminosity.
Our model SF efficiencies $\epsilon$  at T=13.5 Gyr, 
also displayed in  Fig. 6 (lower panel, {\it solid curves}), 
show no dependence on mass or colour.
They also reproduce the observed dispersion, the more compact discs (smaller $\lambda$)
being the more efficient in turning their gas into stars.

The upper and lower panels of Fig. 6, when combined, point to an important
conclusion: {\it since the SF efficiency is independent of galactic mass (or luminosity),
the fact that low mass galaxies have larger gas fractions today may only be due
to their smaller ages}.
This is the most straightforward interpretation, independent of any theoretical
considerations (except for the implicit assumption that the SF efficiency has
remained $\sim$constant during the galaxy's history).
This conclusion is corroborated by an independent observable, namely that
low mass galaxies are, in general, bluer than more massive  ones. Notice that
the latter observable concerns also elliptical galaxies, but the well known problem
of the age-metallicity degeneracy does not allow to conclude in that case.
In the case of spirals, the situation is even worse in principle, because colours may be
affected by the presence of dust (presumably more abundant in massive spirals).
Because of this complication, the observed gas fraction (smaller in large spirals) 
is not sufficient in itself to lift the age-metallicity degeneracy. 
However, when combined with the fact that the SF efficiency is independent of
galactic mass (as argued here), the degeneracy is lift, and the aforementionned
conclusion is naturally obtained.

Before turning to a more detailed discussion of our findings, we would like to 
point out that similar results are obtained in other recent works.
For instance, McGaugh and de Blok (1997) find a clear trend between the gas fraction
and the $B-$magnitude of their galaxy sample, that they describe by the
relation $\sigma_{gas}$=0.12 ($M_B$+23). This relationship is shown in 
Fig. 7 (upper panel, {\it dashed line}), along with the McGaugh and de Blok (1997)
data for normal spirals and our data for normal spirals.
It can be seen that there is a very good agreement between the two data sets.
We notice that 
our models fit the observations of McGaugh and de Blok (1997) 
fairly well (see Fig. 13 of Boissier and Prantzos  2000).

In the lower panel of Fig. 7 we show the SF efficiency, both for our data set and 
the one of Kennicutt (1998b). Notice that Kennicutt (1998b) gives 
the {\it average} SFR surface density $\langle\Psi\rangle$ and {\it average} gas surface density 
$\langle\Sigma_{gas}\rangle$ of normal spirals,
i.e. the integrated quantities are divided by the disc surface aerea (within the
optical radius). Obviously, the ratio 
$\langle\Psi\rangle/\langle\Sigma_{gas}\rangle$ gives the overall SF efficiency
$\epsilon$ (since the disc aerea cancels out). As can be seen in Fig. 7,
Kennicutt's  values of the SF efficiency are slightly larger than ours
(by a factor of $\sim2$), and have a smaller dispersion in the adopted logarithmic
scale. Taking into account the various uncertainties in estimating the SFR
from the data (see Sec. 3), such a discrepancy between Kennicutt's results and ours is
not unexpected.
But the important point is that Kennicutt's values
are also independent of the galaxy 
$B-$luminosity and, by virtue of the Tully-Fisher relation, on the
galaxy's mass.

\begin{figure}
\psfig{file=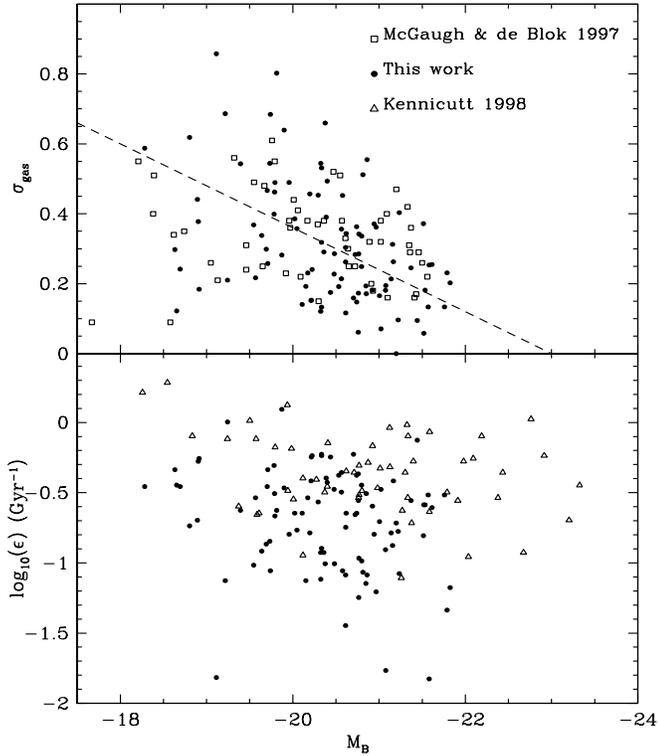,height=11cm,width=0.5\textwidth}
\caption{ 
Gas fraction ($\sigma_{gas}=M_{g}/M_{T}$,
{\it upper panels}) and global 
Star Formation Efficiency ($\epsilon=\Psi/M_{g}$, {\it lower panels}) 
vs $B-$band magnitude. Our data are shown by {\it filled symbols} in both panels.
They are in agreement with 
data from other surveys (McGaugh and de Blok 1997, {\it squares} in the
upper panel; and Kennicutt 1998b, {\it triangles} in the lower panel).
The same trends as in Fig. 6 are also obtained here
(i.e. gas fraction decreasing with galaxy's
luminosity and Star Formation efficiency independent of luminosity). The 
{\it dashed curve} in the upper panel is a fit to the McGaugh and DeBlok (1997)
data: $\sigma_{gas}$=0.12($M_B$+23).
}
\end{figure}

The anti-correlation between gas fraction and luminosity in spirals was
noticed by several authors (e.g. Gavazzi et al. 1996, McGaugh and de Blok 1997,
Bell and de Jong 2000, Boselli et al. 2000, etc.). Based on two independent samples, we showed here
that the SF efficiency of spirals is independent of their mass. The two findings
combined, point to small discs being younger, on average, than massive ones.
Our model, presented in Sec. 3.2, nicely explains these features (and several
other, presented in papers II and III). However, one may argue that the
complex interdependence between the adopted infall and SFR prescriptions
makes a straightforward interpretation difficult; he/she may also
argue that other types of models could also account for the observations and
give a different interpreration for the same data (e.g. by invoking outflows).
For that reason, we discuss in the next section this issue on the basis
of simple analytical models of galactic chemical evolution and we make a very rough 
evaluation of the ``ages'' of the galaxies in our sample.
Our purpose is not to derive the exact ages of the galaxies, but rather to have
an order of magnitude estimate and, in particular, to
check whether there is any {\it trend} of the derived ages with galactic mass.

\section{Galactic Ages}

In the framework of simple models of galactic chemical evolution adopting
the Instantaneous Recycling Appproximation (IRA), one may
obtain analytical solution for various quantities. In particular, provided
that the Star Formation Rate $\Psi$ is proportional to the gas mass $M_g$,
i.e.
\begin{equation}
\Psi=\epsilon M_{g} \label{equagas}
\end{equation}
one may obtain a relationship between the gas fraction $\sigma_{gas}$ and time T
(assuming that the SF efficiency $\epsilon$ is constant in time).
The form of this relationship depends on further assumptions about the
evolution of the system, i.e. on the possibility of allowing for gas 
flows in or outside the ``box'' (e.g. Pagel 1997).

Assuming that the galaxies of our sample have evolved as simple, homogeneous,
``boxes'', we consider three possibilities:
a ``closed box'' (all the gas is present from the very beginning),
an ``infall'' model (where gas mass is continuously added to the system) and
an ``outflow'' model (with gas continuously leaving the system).
In the cases of gaseous flows, further assumptions about the corresponding
flow rates are required in order to obtain analytical solutions.
More specifically:

\noindent a) {\it Closed Box}: In that case, we have
\begin{equation}
\sigma_{gas} \ = \ {\rm exp}[-(1-R) \, \epsilon \, T]
\end{equation}
where the return fraction $R$ accounts for the gas returned by stars
to the interstellar medium; for the IMF of Kroupa et al. (1993) adopted here
we have $R\sim$0.32.

\noindent b) {\it Infall}: An analytical solution is easily
obtained  if it is assumed that
the infall rate just balances the gas depletion due to star formation.
As we have shown in Paper II, with detailed numerical models reproducing
a large body of observational data, this situation describes rather well
the largest period in the lifetime of most spiral galaxies.
In that case, we have:
\begin{equation}
\sigma_{gas} \ = \ [1+\epsilon(1-R)T]^{-1}
\end{equation}

\noindent c) {\it Outflow}: Analytical solutions may be obtained by assuming that the
outflow rate is proportional to the SFR: $f_{out}=\gamma\Psi$.
In that case, we have:
\begin{equation}
\sigma_{gas}=\frac{R-1-\gamma}{ (R-1) {\rm exp}[- \epsilon (R-1-\gamma) T] - \gamma}
%\sigma_{g} \ = \ exp[-\epsilon (1-R+\gamma)T]
\end{equation}
We shall assume here that the
outflow rate is equal to the SFR, i.e. $\gamma$=1 
(since, for higher outflow rates, we obtain ridiculously low galactic ages). 

\begin{figure}
\psfig{file=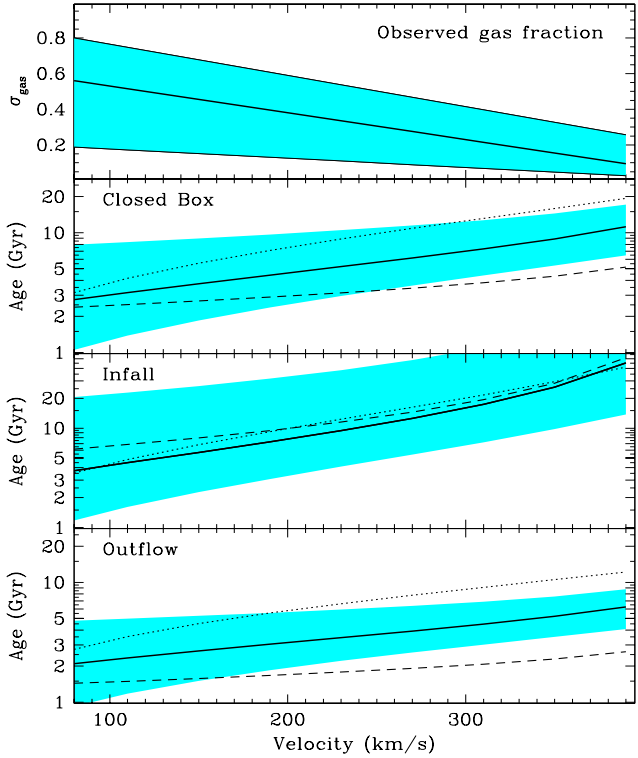,height=11cm,width=0.5\textwidth}
\caption{
{\it Upper panel}: Observed gas fraction vs. galaxy's rotational velocity;
the three curves represent, schematically, the mean trend ({\it thick curve})
and the upper and lower bounds of the observed gas fractions in Fig. 6.
Given the Star Formation efficiency $\epsilon$,
the shaded aerea inside the upper and lower curves may be used to derive 
approximate ages for the galaxies, in the
framework of simple analytical models for galactic chemical evolution, using
IRA (Instantaneous Recycling Approximation). This is done in the three
next panels, for a Closed Box, a model with Infall (such as the gas mass remains
constant in time) and a model with Outflow (with an outflow rate equal to the
Star Formation Rate). In each of those panels, the {\it thick curve} corresponds
to the mean gas fraction of the upper panel and a SF efficiency 
$\epsilon$=0.3 Gyr$^{-1}$ (the mean value for the efficiencies in Fig. 6).
The {\it shaded aerea} is obtained by keeping the same   SF efficiency, but
considering the upper and lower bounds for the observed gas fraction in the upper panel
(lower gas fractions lead to larger ages for a given rotational velocity).
The {\it dotted curves} are obtained with the low gas fractions combined to
a high SF efficiency ($\epsilon$=1 Gyr$^{-1}$, a reasonable upper bound for the
observed $\epsilon$ values in Fig. 6). 
Finally, the {\it dashed curves} are obtained with 
the high gas fractions combined to
a low SF efficiency ($\epsilon$=0.1 Gyr$^{-1}$, a reasonable lower bound for the
observed $\epsilon$ values in Fig. 6).
}
\end{figure}

For each of the three scenarios, we shall consider five combinations
between the observed gas fraction and SF efficiency, in order to derive
galactic age T through Eqs. 6, 7 and 8.
For the gas fraction vs. rotational velocity, 
we adopt the three curves of the upper panel of Fig. 8,
corresponding roughly to the mean trend and the upper and lower bounds
of the observations, respectively.
For each of those three curves, we derive the corresponding galactic ages by
using a SF efficiency $\epsilon$=0.3 Gyr$^{-1}$ (the mean value of the
observations in Fig. 6 and 7); in that way, we obtain the {\it shaded} regions 
in Fig. 8, with the thick curve corresponding to the mean trend, the largest
ages to the lowest gas fractions and vice versa.
We perform two more calculations. For the first, we adopt the high gas fractions
(uppermost curve in the upper panel of Fig. 8), combined to a low
SF efficiency $\epsilon$=0.1 Gyr$^{-1}$ (a reasonable lower bound for the
observations of Fig. 6 and 7); in that way we obtain the {\it dotted curves} in 
Fig. 8. Finally, we adopt the low gas fractions
(lower curve in the upper panel of Fig. 8), combined to a high
SF efficiency $\epsilon$=1 Gyr$^{-1}$ (a reasonable upper bound for the
observations of Fig. 6 and 7); in that way we obtain the {\it dashed curves} in 
Fig. 8.

An inspection of the results, plotted in the three lower panels of Fig. 8,
shows that the derived ``effective'' galactic age is a monotonic function of
rotational velocity $V_C$ in all cases.
The absolute ages depend, of course, on the adopted gas fractions, SF efficiencies and
assumptions about gaseous flows. For a given gas fraction and SF efficiency, the infall 
model leads to the largest ages; since the gas is constantly replenished, it takes more
time to attain a given gas fraction than in a closed box (starting with $\sigma_{gas}$=1).
Also, the outflow model produces the lowest ages; since part of the gas is
constantly removed, it takes less time to reach a given gas fraction than in a closed
box. The ages derived in the framework of the outflow model are, in general,
too low: they are lower than 6 Gyr for all galaxies with $V_C<$200 km/s. 
It transpires that
galaxies in this velocity range should not have suffered extensive mass losses, i.e.
they must not have lost during their lifetime an amount of gas as important
as their stellar content (since we adopted $f_{out}=\Psi$ here).

The mean age values in the case of the infall model are 4-5 Gyr for the small
discs (in the $V_C\sim$100 km/s range) and $\sim$10 Gyr for $V_C\sim$220 km/s.
For discs with $V_C>$300 km/s, extremely large ages ($>$15 Gyr) are found; however,
the approximation of evolution at constant gas mass is certainly not valid in that case
(see Fig. 4 in Boissier and Prantzos 2000).
We notice that Bell and de Jong (2000), on the basis of a different sample and
with a completely different method (based on a photometric estimate of the ages)
find an ``effective age'' of $\sim$10 Gyr for the more luminous galaxies ($M_K\sim$-26)
and $\sim$6 Gyr for the less luminous ones ($M_K\sim$-20); the galaxies of
our sample span also this luminosity range, and its interesting to see that
similar ages are found with completely independent methods.

The results obtained in this section confirm the suggestion of Sec. 3.2:
low mass discs are, on average, younger than massive ones. The only way
to reverse this trend is by assuming that the SF efficiency is strongly correlated
with disc mass. Despite the large scatter in the observational data, it is clear that 
such a correlation does not exist, at least at the present time; and,
since the observed galaxies span a large range in masses and metallicities, 
it does not seem plausible that such a correlation ever existed in the past.
In the case of outflow models, another possibility would be to consider that the more
massive discs suffered more important mass losses (leading to low gas fractions
without having to invoke large ages). However, such a hypothesis is incompatible to the
fact that massive discs have deeper potential wells and are less prone to outflows than
low mass ones.

\section{Summary}

In this work we investigate the properties of the star formation efficiency of 
spiral galaxies and study the implications for their evolution. We use a large homogeneous 
sample of disc galaxies, for which we measure gaseous mass $M_g$ (HI), 
rotational velocity 
$V_C$, star formation rates $\Psi$ and luminosities in the $B$ and $H$ bands. We are
then able to derive the corresponding gas fractions $\sigma_{gas}$ and SF efficiencies 
$\epsilon=\Psi/M_g$ as a function of $V_C$, $H-$luminosity and $B-H$ colour index.
We find that the gas fraction is correlated to $V_C$, $H-$luminosity and $B-H$, in the
sense that more massive, luminous and redder discs have smaller gas fractions.
Previous work by McGaugh and de Blok (1997) reached similar conclusions.
The main finding of this work is that the SF efficiency does not correlate with any of
the galaxy properties; despite a rather large dispersion (within a factor of $\sim$10),
the observed $\epsilon$ is independent of $V_C$, $L_H$ or $B-H$.

We interpret our data in the framework of detailed models of galactic chemical and
spectrophotometric evolution, utilising metallicity dependent stellar lifetimes, yields,
tracks and spectra. These models are calibrated on the Milky Way disc (paper I) and use
radially dependent star formation rates, which reproduce observed gradients (paper III).
They are extended to other spirals  in the framework of Cold Dark Matter scenarios for 
galaxy formation, and are described by two parameters: rotational velocity $V_C$ and spin
parameter $\lambda$. As in our previous work (paper II), we find good agreement with
the observations, provided a crucial assumption is made: massive discs are formed earlier
than less massive ones. With this assumption our models reproduce the observed 
trends of gas fractions and SF efficiencies vs. $V_C$, while variations due to $\lambda$
account for the observed dispersion in both cases.

It is important to notice that
the dependence of age on galactic mass that we find is not due to any mass-dependent
SF efficiency, only to disc formation timescales; in our models this
is achieved by varying the infall timescales. Both observations and models suggest
that the SF efficiency is independent of galaxy properties. Since the observed galaxies
cover a wide range of masses, colours and metallicities, there is no reason
to suppose that the SF efficiency was different in the past. The adopted SFR
prescription in our models also results in a very slowly 
varying SF efficiency with time.

When the observed relations of gas fraction vs $V_C$ and SF efficiency vs $V_C$
are considered in combination, they
convincingly suggest that low mass discs are, on average, younger than more massive ones;
this conclusion is independent of any model and the only assumption is that the SF 
efficiency is constant in time (a quite plausible assumption, as argued above).
In the framework of simple analytical models of galactic chemical evolution, we evaluate
the ``effective ages'' for the galaxies of our sample, using the observationally 
derived values of
$\sigma_{gas}$ and $\epsilon$. We find that even models with modest outflows (with ejected
masses equal to the stellar ones) lead to ridiculously low values for the galaxy ages;
our conclusion is that galaxies in the range $V_C\sim$80-400 km/s have not suffered 
extensive mass losses. Closed box models and infall models lead to
more plausible values for the effective ages. In particular, infall models
lead to ages of $\sim$4-5 Gyr for discs of $V_C\sim$100 km/s and $\sim$8-10 Gyr for
$V_C\sim$200 km/s. These ``chemically derived'' ages are in fair agreement with those
derived on the basis of our more sophisticated numerical models, which
fit  a much larger body of observational data for low redshift spirals.
Most importantly, 
they are also in fair agreement with the ``photometric'' 
ages derived in a completely independent
way and with a different sample by Bell and de Jong (2000).

In summary, our data, taken at face value, suggest that {\it the bulk of stars
in more massive discs
are older than in less massive ones}. This is supported by our detailed
numerical models of galactic chemical and photometric evolution,
but also by recent, independent, analysis (Boselli et al. 2000).
We notice that Bell and deJong (2000) conclude that it is local surface density
that mainly drives the star formation history, while mass plays a less
important role. Our study suggests that mass is the main factor, while
local surface density plays only a minor role (through the spin parameter
$\lambda$: lower $\lambda$ values lead to higher local surface densities
for a given rotational velocity $V_C$).

We notice that this picture is hardly compatible with the currently popular
``paradigm'' of hierarchical galaxy formation, which holds that
large discs are formed by merging of small units at relatively late epochs.
If this were the case, massive discs should have large SF efficiencies, in order
to have their gas fractions reduced to lower levels than  their less massive
counterparts. However, such an enhanced SF efficiency is not supported by observations.

\label{lastpage}

\end{document}